\documentclass[journal abbreviation, manuscript]{copernicus}

\usepackage{subcaption}
\usepackage{upquote}
\nolinenumbers

\begin{document}

\title{On the prediction of underwater aerodynamic noise of offshore wind turbines}


\Author[1][l.botero@upm.es]{Laura Botero-Bolívar}{} 
\Author[1]{Oscar A. Marino Sánchez}{}
\Author[1]{Martín de Frutos}{}
\Author[1]{Esteban Ferrer}{}

\affil[]{ETSIAE-UPM - School of Aeronautics, Universidad Politécnica de Madrid, Plaza Cardenal Cisneros 3, Madrid}




\runningtitle{TEXT}

\runningauthor{TEXT}



\firstpage{1}

\maketitle

\begin{abstract}
The growing demand for offshore wind energy has led to a significant increase in wind turbine size and to the development of large-scale wind farms, often comprising 100 to 150 turbines. However, the environmental impact of underwater noise emissions remains largely unaddressed. This paper quantifies, for the first time, the underwater aerodynamic noise footprint of three large offshore turbines (5 MW, 10 MW, and 22 MW) and wind farms composed of these turbines. We propose a novel methodology that integrates validated wind turbine noise prediction techniques with plane wave propagation theory in different media, enabling turbine designers to predict and mitigate underwater noise emissions. Our results confirm that aerodynamic noise from offshore wind farms presents a potential environmental challenge, with negative effects on marine life. Addressing this issue is crucial to ensuring the sustainable expansion of offshore wind energy.
\end{abstract}


\introduction  
Anthropogenic underwater noise significantly impacts marine ecosystems, posing the natural equilibrium among different species~\cite{Gotz2009}. Marine mammals are especially vulnerable to noise since they rely on sound to communicate, navigate, reproduce, feed, among other sensory purposes. A commonly cited problem concerning underwater noise is the masking problem~\citep{Erbe2016}, where anthropogenic noise can mask sound naturally present in marine environments, leading to alteration of behavior, reduction of communication ranges, foraging, predator and habitat avoidance~\citep{StöberandThomsen, Daly2012}. Furthermore, the propagation of sound in the sea is enhanced by the higher speed of sound in water than in air ($1480$~m/s for water vs. $343$~m/s for air) and the lower sound attenuation in water ($0.1$~dB/km in seawater vs. $5$~dB/km for air at $1$~kHz). These physical factors, together with the channeling of sound in shallow waters, make anthropogenic noise more critical underwater. For example, \citet{Tougaard2020} measured underwater noise $20$~km away from a small offshore wind farm composed of only 16 wind turbines. Therefore, predicting underwater noise produced by offshore energy devices is paramount to guarantee sustainable exploitation of energy sources. 

Wind turbine design for offshore environments has primarily focused on maximizing energy production (e.g.,~\cite{Akhtar2024,doi:10.1126/sciadv.aax9571,DESALEGN2023100691}), with limited attention given to the acoustic footprint and its environmental impact. Specifically, the underwater aerodynamic noise generated by offshore wind turbines has not yet been quantified, which is the focus of this work.

Aerodynamic noise of wind turbines is caused by the interaction between the turbulent wind and the turbine blades. Thus, it depends on the operational conditions and size of the wind turbines - the overall sound pressure level of a wind turbine scales with the fifth power with the wind turbine rotor diameter. In recent years, the size of offshore wind turbines has increased steadily in response to the growing demand for clean energy production. For example, the offshore wind turbine proposed within the IEA Task 55—REFWIND would produce $22$~MW power with a rotor diameter of $284$~m~\citep{IEA22MW} and the recently announced MySE $22$~MW offshore turbine from Mingyang Smart Energy with a rotor of $310+$ meters~\citep{Mingyang22mw}. In addition, we face a rapid increase in the number of turbines gathered in farms. Today, the largest offshore wind farms (e.g., London Array, Gemini, Hornsea Project One and Two~\citep{orsted_hornsea1}) include more than 150 turbines. Simple acoustics shows that the noise produced by a wind farm of $N$ wind turbines scales with the factor $20\log_{10} (N)$. The combination of increasingly large turbines gathered in farms with hundreds of turbines, combined with the negative impact of underwater noise, motivates this research: Can aerodynamic noise from offshore wind turbines affect marine life? To quantify this issue, we develop a new approach which combines wind turbine noise prediction techniques with wave theory to calculate the effective noise that penetrates underwater (due to the change of media).

Wind turbine noise may be classified as mechanical and aerodynamic acoustic noise. The first type has a defined tonal character and is produced by mechanical components such as the gearbox and bearings (and/or generator or cooling systems) located within the device nacelle and may be controlled/minimized by appropriate insulation of the nacelle. The second type is more complex and is caused by the interaction of the blades moving through the air. Previous studies on offshore wind turbine noise have only considered mechanical noise as it is directly propagated into the water through the vibrating tower (or platform)~\citep{Tougaard2020,Hooper2003, madsen2006,Marini2017,Thomsen2006,Tougaard2009}. 

A common justification for ignoring the aerodynamic noise (produced by the rotating blades) of offshore wind turbines is based on Snell's law~\citep{Chapman1990}, which states that only one portion of the noise produced in the air propagates into the water. For air-water interfaces, Snell states that only sound waves within a cone of 13\textsuperscript{o} angle with respect to the air-water interface normal vector can propagate into the water, see figure \ref{fig:noise_prop}. This fact considerably limits the propagation of airborne noise sources into the water. Furthermore, the higher acoustic impedance of water compared to air (i.e., the acoustic impedance of water is $3600$ times higher than the air's) leads to a high attenuation of the sound waves when entering water. 
The main contribution of this work is to show that despite this high attenuation, underwater noise is still potentially dangerous to marine animals. We couple wind turbine aerodynamic noise prediction methods with plane wave theory and Snell's law to predict wind turbine noise underwater. This contribution is a breakthrough for manufacturers and environmentalists, as it allows to consider noise generation in turbine design. Furthermore, The method is a fast turn-around method that can be used in low-fidelity models, but it can be easily combined with high-fidelity models to characterize the noise source or noise propagation. 

Using the proposed approach, we will quantify how the underwater footprint of large offshore turbines can affect several marine species. We compute the aerodynamic noise of $5$~MW, $10$~MW, and $22$~MW offshore wind turbines and compare the underwater transmitted sound to the hearing thresholds of many marine animals. We confirm that these emissions are an environmental problem that is exacerbated when large offshore farms with hundreds of turbines are built.

The remainder of the paper is organized as follows. Section~\ref{sec2} presents the methodology used in the research, including wind turbine noise prediction, air-water interface modeling, and the characteristics of wind turbine models. Section~\ref{sec3} addresses the results and discussion. Finally, Section~\ref{sec4} shows the main conclusions of this research. 

\section{Methodology}\label{sec2}

\subsection{Wind turbine noise predictions}
\begin{figure}[h!]
\centering
\includegraphics[width=0.6\textwidth]{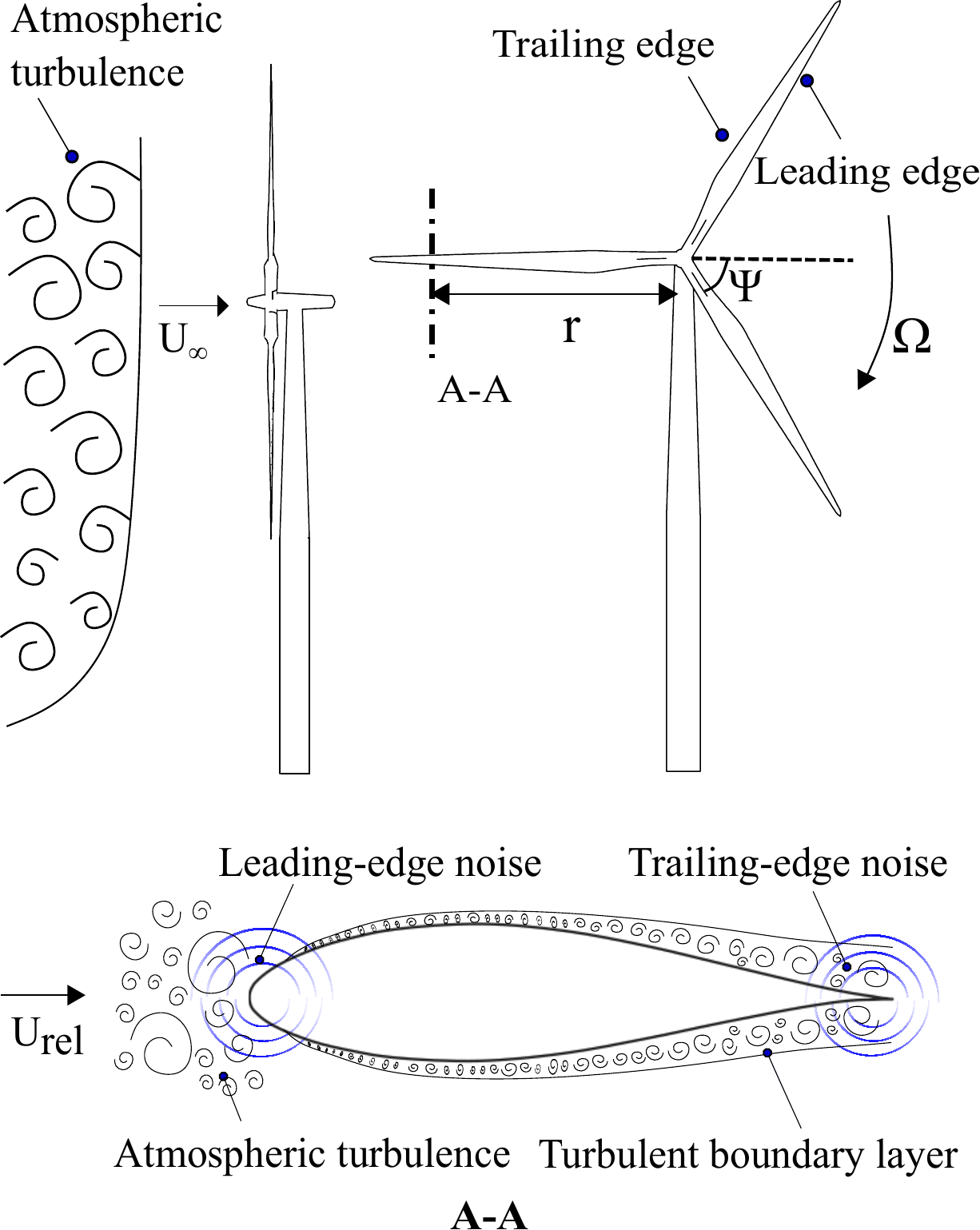}
\caption{Aerodynamic noise sources for an offshore wind turbine. }\label{fig:Wt_noise}
\end{figure}

Aerodynamic noise, caused by the interaction of the flow with the structure, is the main source of noise of modern wind turbines~\cite{Oerlemans2007}. \autoref{fig:Wt_noise} sketches the typical aerodynamic environment and the sources of noise of an offshore wind turbine. The atmospheric turbulence interacts with the leading edge of the rotating blades, causing a low-frequency noise, known as leading edge (LE) noise. Additionally, the turbulent boundary layer on the blades that interacts with the finite trailing edge causes mid- to high-frequency noise, referred to as trailing edge (TE) noise. Overall, the wide range of turbulent scales -- from hundreds of meters in atmospheric flow to millimeters in boundary layer flow -- encountered by the wind turbine blades cause aerodynamic noise to exhibit a broadband nature, covering a wide range of frequencies. This is particularly critical for marine environments, as it can affect a variety of marine species with different hearing thresholds.

To compute the total aerodynamic noise of the wind turbine, we follow the method proposed by Schlinker-Amiet~\cite{schlinker1981} for rotatory noise sources. We consider the strip theory approach, where the blade is divided into $n$ segments. Each segment is considered as a 2D airfoil (as shown in the A-A cut in~\autoref{fig:Wt_noise}). For each segment, leading- and trailing-edge noise ($S_{pp|\mathrm{LE}}$ and $S_{pp|\mathrm{TE}}$) are calculated as uncorrelated noise sources, such as: $S_{pp|seg} = S_{pp|\mathrm{LE}} + \textcolor{black}{S_{pp|\mathrm{TE}}}$. 

Leading- and trailing-edge noise (LE and TE) are predicted using Amiet's theory~\citep{amiet1975LE, amiet1976TE} and the extension of \citet{roger2005back} to consider the back-scattering effect caused by airfoils of finite chords. The blade is divided into segments that are more refined close to the blade tip, which is the part that generates most of the noise. An initial sinusoidal distribution of the location of the segments is proposed. The sinusoidal distribution is obtained by the horizontal coordinate of a point located in a semicircle of a diameter equal to the rotor radius (neglecting the inner part of the blades that consist of cylinders). The angle between the points in the radial axis was constant. After an iterative process to ensure that the aspect ratio ($AR$), defined as the span-chord ratio of the blade section, is larger than three, the radial position and the span- and chord-length distribution are obtained. The $AR \geq 3$ condition is adopted to satisfy the far-field condition assumed in Amiet's theory.

The von Kármán model~\citep{vonKarman1948} calculates the inflow turbulence spectrum used as input for predicting LE noise and an extension of TNO-Blake model~\citep{Stalnov2016} computes the wall pressure spectrum to calculate trailing-edge noise. The boundary layer characteristics used as input in the TNO-Blake model are computed by XFOIL~\citep{Drela_XFOIL} using the flow conditions (angle of attack, $\alpha$ and relative velocity, $U_\mathrm{rel}$) obtained with the blade element momentum theory (BEMT). The transition for XFOIL simulations was fixed at 5\% of the chord. BEMT solutions are obtained using the open-source code OpenFAST~\citep{OPENFAST}, which includes the Prandlt tip and root loss correction factors and the Pitt / Peters skewed wake correction model.

The relative motion of the segment with respect to a fixed observer, due to the rotation of the blades, induces a delay between the noise emission and the location of the observer. This delay is quantified by a Doppler factor ($(\omega_e/\omega)^2$) that is incorporated in the prediction method as the squared ratio between the emitted frequency ($\omega_e$) and the received frequency ($\omega$). Subsequently, the total blade noise is calculated as the sum of all segments, computed at every azimuth angle ($\Psi$ in~\autoref{fig:Wt_noise}):
\begin{equation}\label{eq:spp_blade}
	S_{pp|\mathrm{blade}}(\omega,\Psi) = \sum_1^n S_{pp|\mathrm{seg}}(\omega_e,\Psi) (\omega_e/\omega)^2.
\end{equation}

The average noise produced by the wind turbine in one rotation ($ S_{pp|\mathrm{WT}}(\omega)$) is then calculated as: 
\begin{equation}\label{eq:Spp_WT}
	S_{pp|\mathrm{WT}}(\omega)  = \frac{B}{2\pi} \int_0^{2\pi} S_{pp|\mathrm{blade}}(\omega,\Psi) \mathrm{d} \Psi,
\end{equation}
where $B$ is the number of blades. More information on the noise prediction methodology can be found in \citet{boterobolivar2024}.

To compute the noise generated by a wind farm, we assume that each turbine acts as an uncorrelated noise source with equal intensity and is positioned at the same relative location with respect to the observer, i.e., adding a factor of $10\log_{10} N$, where $N$ is the number of wind turbines. In an actual scenario, a single global observer for the wind farm would be at a different relative location for each wind turbine within the farm. This introduces a significant dependence of the noise produced by the wind farm on its layout, which is beyond the scope of this work. 

\subsection{Air-water interface modeling}\label{sec:air_water_trans}
\begin{figure}[h]
\centering
\includegraphics[width=0.95\textwidth]{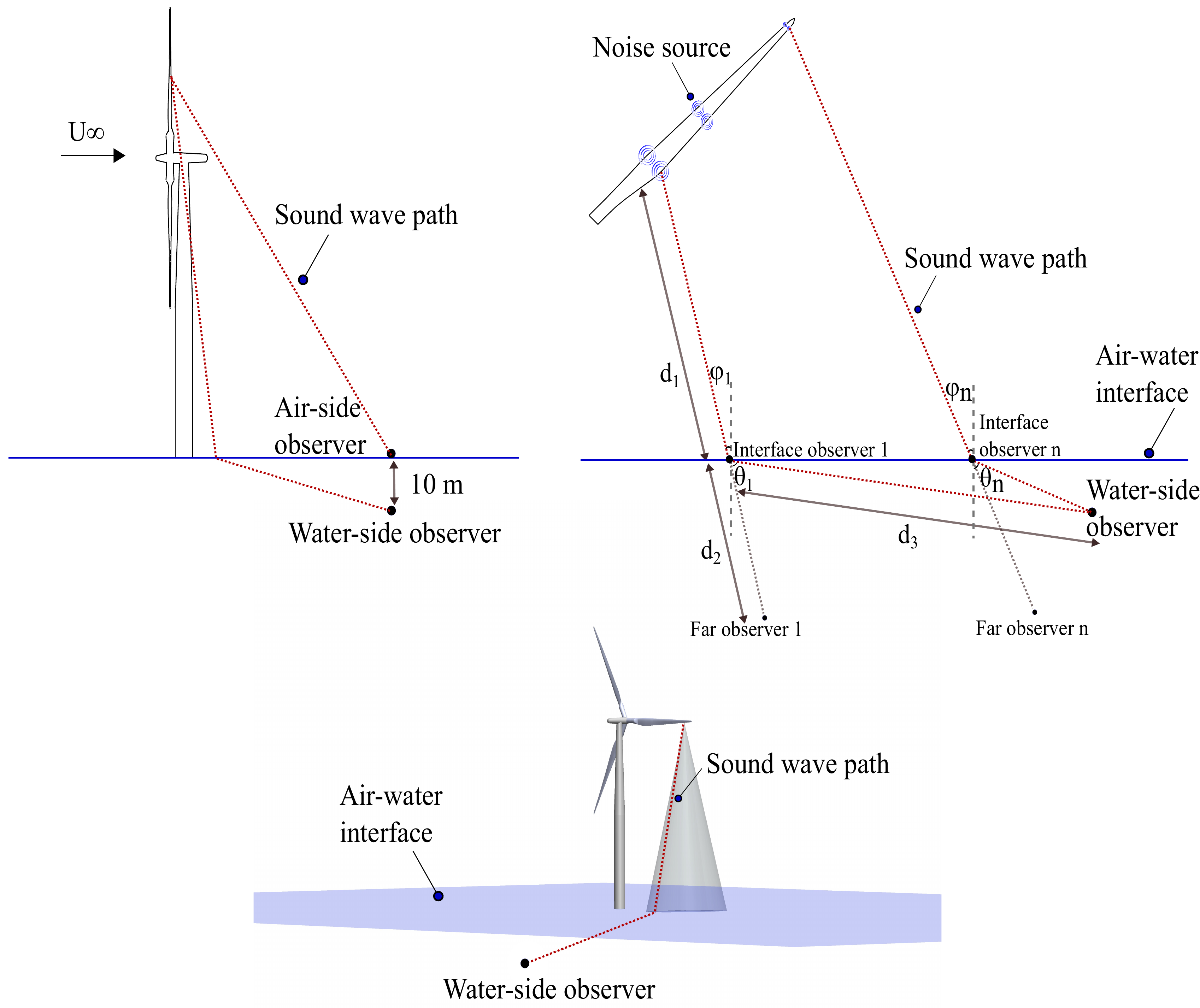}
\vspace{5mm}
\caption{Sketches of wind turbine noise trajectories for an Air-side and a Water-side observers.}\label{fig:noise_prop}
\end{figure}

The noise is calculated at two observers downstream of the turbine, as shown in~\autoref{fig:noise_prop}~left. The Air-side observer is located exactly at the air-water interface and the Water-side observer is located at 10~m depth from the air-water interface, both at a specific downstream position that is defined for each case analyzed in the results. Here, we pay particular attention to the propagation of noise underwater, therefore, additional steps need to be considered for the Water-side observer compared to the Air-side observer.

For the Air-side observer, the far-field noise is calculated directly at the observer, following the standard procedure of wind turbine noise prediction~\cite{boterobolivar2024}. However, for the Water-side observer, the far-field noise cannot be computed directly at the observer because of the change of media that causes refraction and attenuation of the sound waves. This means that for each noise source (LE and TE of each segment at each azimuth location) only the direction of the sound waves that refracted at the air-water interface reaches the Water-side observer is considered and the sound pressure level is attenuated at the air-water interface. The rest of this subsection will address the details for the computation of the underwater noise. Furthermore, the algorithm with the methodology to compute the far-field noise spectrum at any Water-side observer is shown in \ref{app:algoritmo}.

The refraction of the sound waves due to the change of media is computed following the Snell's law that establishes:
\begin{equation}\label{eq:snell}
    c_w \sin{\phi} = c_a \sin{\theta};
\end{equation}
where $c_w$ and $c_a$ are the speed of sound in water and air, respectively, and $\phi$ and $\theta$ are the angles formed by the sound wave and an axis perpendicular to the air-water interface in the air and the water, respectively, as shown in~\autoref{fig:noise_prop}~right. In this research, we consider a flat air-water interface. $\phi$ is calculated for each blade segment at each azimuth location, considering the location of the noise source (segment noise) and the Water-side observer. With $\phi$ defined, an \textit{interface observer} is defined for each noise source, which is located in the intersection of the vector that connects the air-water interface with the noise source (following the angle $\phi$) and the air-water interface, as shown in \autoref{fig:noise_prop}~right. The far-field noise of each segment is then first calculated at the interface observer following the noise prediction approach for a single medium (i.e., Amiet's theory for LE and TE noise). Subsequently, the noise is propagated from this interface observer to the Water-side observer through the water media. The blade noise, i.e., the sum of all the segments, and the full wind turbine noise over the rotation, is calculated at the water-side observer.

According to Snell's law, for an air-water interface, the maximum value of $\phi$ is $\approx$ 13\textdegree, which locates the interface observer very close to the noise source; thus, the far-field condition assumed in Amiet's theory is not satisfied. To overcome this limitation, LE and TE noise are precomputed at a \textit{far observer} (see \autoref{fig:noise_prop}~right) which is located in the same direction with respect to the blade segment as the interface observer but at an arbitrarily long distance from the noise source ($d_1+d_2$ in \autoref{fig:noise_prop}). Later, segment noise is rescaled to the interface observer considering a spherical propagation of the noise, i.e., following the expression $10\log_{10} \left(1/r^2\right)$, where $r$ is the distance between the interface observer and the far observer, i.e., $d_2$ in~\autoref{fig:noise_prop}. 

The transmission loss of the acoustic waves across the air-water interface is calculated following plane wave theory~\cite{Chapman1992}:
\begin{equation}\label{eq:tl}
  TL_{1 \rightarrow 2} = \frac{4gz}{(1+gz)^2};  
\end{equation}
where $g$ is the ratio between the densities of medium 1 and 2, and $z$ is the ratio between the speeds of sound. For the air-water interface, the factor $gz \approx 1/3600$, which gives an intensity transmission loss of 29.5~dB.

The propagation from the interface observer to the Water-side observer is carried out using a cylindrical propagation, i.e., following the expression $10\log_{10} \left(1/r\right)$, where $r$ is the distance from the interface observer to the Water-side observer, i.e., $d_3$ in~\autoref{fig:noise_prop}. Cylindrical propagation is appropriate for underwater noise in shallow waters~\citep{erbe2022, Chapman1990}. 

The atmospheric attenuation ($A$) is considered for both, Air-side and Water-side observers, as follows:
\begin{equation}
   A = \alpha_a r; 
\end{equation}
where $\alpha_a$ is the attenuation in dB/m and $r$ is the distance from the noise source to the observer. For the Air-side observer $d$ is the linear distance from the noise source to the observer, and for the Water-side observer, $d$ is the distance from the noise source to the interface observer ($d_1$ in~\autoref{fig:noise_prop}). $\alpha_a$ is calculated as the attenuation of a pure tone sound wave because of traveling through the atmosphere. It is calculated by the standard ANSI/ASA S1.26~\cite{hansen2017, ansi2014}. The atmospheric conditions used in this research for the calculation of atmospheric attenuation are: $T~=~$15~\textdegree C and $T_\mathrm{ref}~=~$20~\textdegree C;  $P~=~$98~kPa and $P_\mathrm{ref}~=~$101.325~kPa; and $h~=~$86\%; where $T$ and $T_\mathrm{ref}$ are the source and reference temperatures, $P$ and $P_\mathrm{ref}$ are the source and reference atmospheric pressures, and $h$ is the relative humidity. 

\subsection{Offshore wind turbine models}
In our study, we consider three offshore wind turbines that span a wide range of geometric and operational conditions, to consider the effect of size on aerodynamic noise and the impact on marine life: the NREL~5~MW~\citep{NREL5MW}, the DTU~10~MW~\citep{DTU10MW}, and the IEA~22~MW~\citep{IEA22MW}. \autoref{tab:WT} summarizes the geometrical details and nominal operational conditions of the turbines. We assume a turbulence intensity of 9\% and an integral length scale of 100~m for all cases since these are typical values for offshore sites~\citep{kale2023, TurkandEmeis}. Those values are used to compute the von Kármán spectrum to predict LE noise. The effect of turbulence intensity is also discussed in the results section. 
\begin{table}[h]
\caption{Geometrical and operational (nominal) conditions of three large offshore wind turbines}\label{tab:WT}%
\begin{tabular}{@{}lccc@{}}
\hline
Characteristic & NREL~5~MW  & DTU~10~MW & IEA~22~MW\\
\hline
Hub height [m] & 90.0   & 119.0 & 170.0 \\
Rotor diameter [m] & 126.0   & 178.4  & 284.0  \\
Nominal wind velocity [m/s]    & 11.4   & 11.4  & 11.78  \\
Rotor angular velocity [rpm]    & 12.1   & 9.6  & 7.1  \\
Blade tip velocity [m/s] & 79.0  & 90.0  & 102.0  \\
Blade Pitch [deg.] & 0.0   & 0.0  & 6.6 \\
\hline
\end{tabular}
\end{table}

\section{Results and Discussion}\label{sec3}
\begin{figure}[h!]
	\centering
		\begin{subfigure}[b]{0.49\textwidth}
		\centering
		\includegraphics[width=\textwidth]{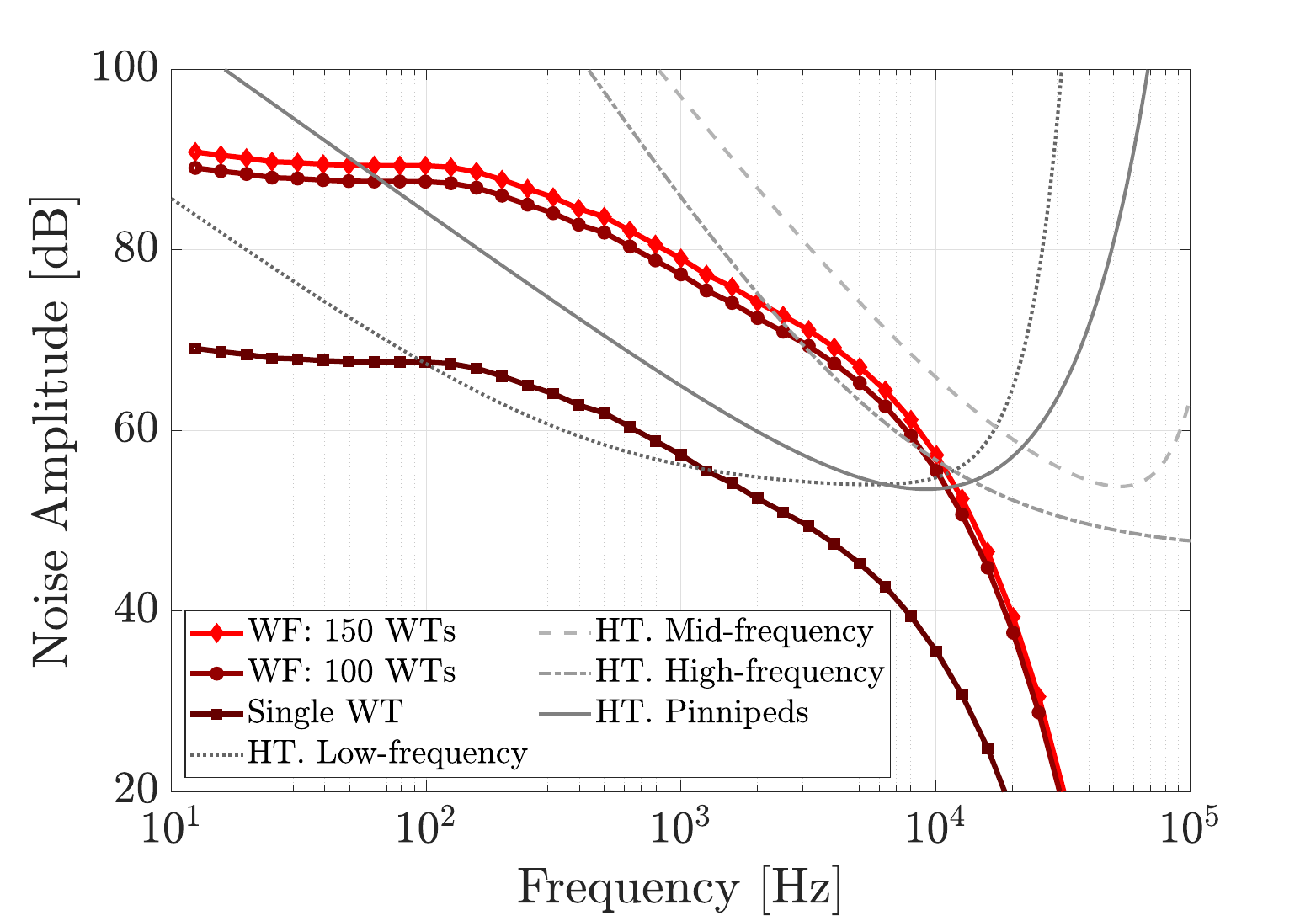}
		\caption{NREL 5~MW wind turbine}
		\label{subfig:5MW}
	\end{subfigure}
	\hfill
	\begin{subfigure}[b]{0.49\textwidth}
		\centering
		\includegraphics[width=\textwidth]{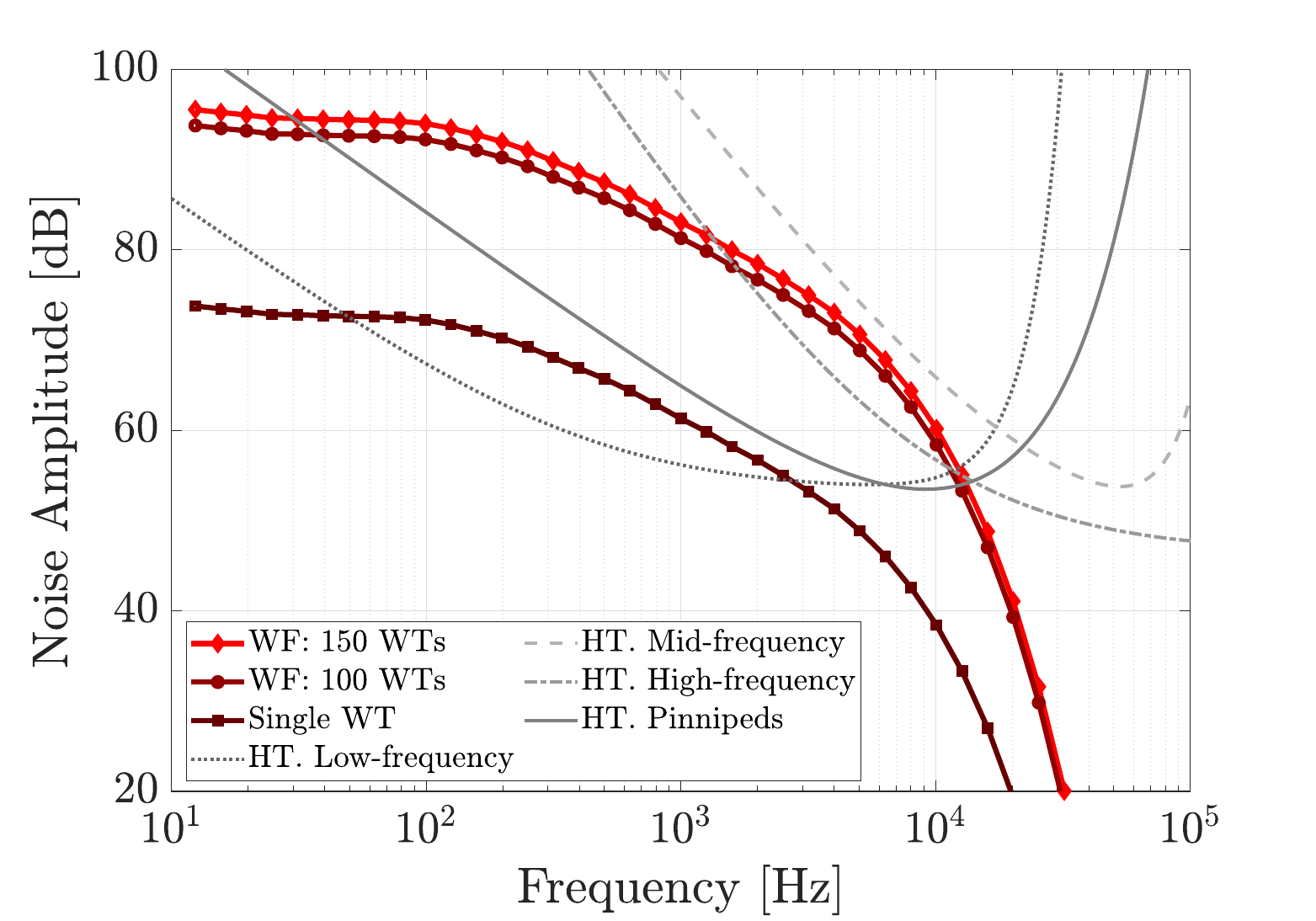}
		\caption{DTU 10~MW wind turbine}
		\label{subfig:10MW}
	\end{subfigure}
 	\begin{subfigure}[b]{0.49\textwidth}
		\centering
		\includegraphics[width=\textwidth]{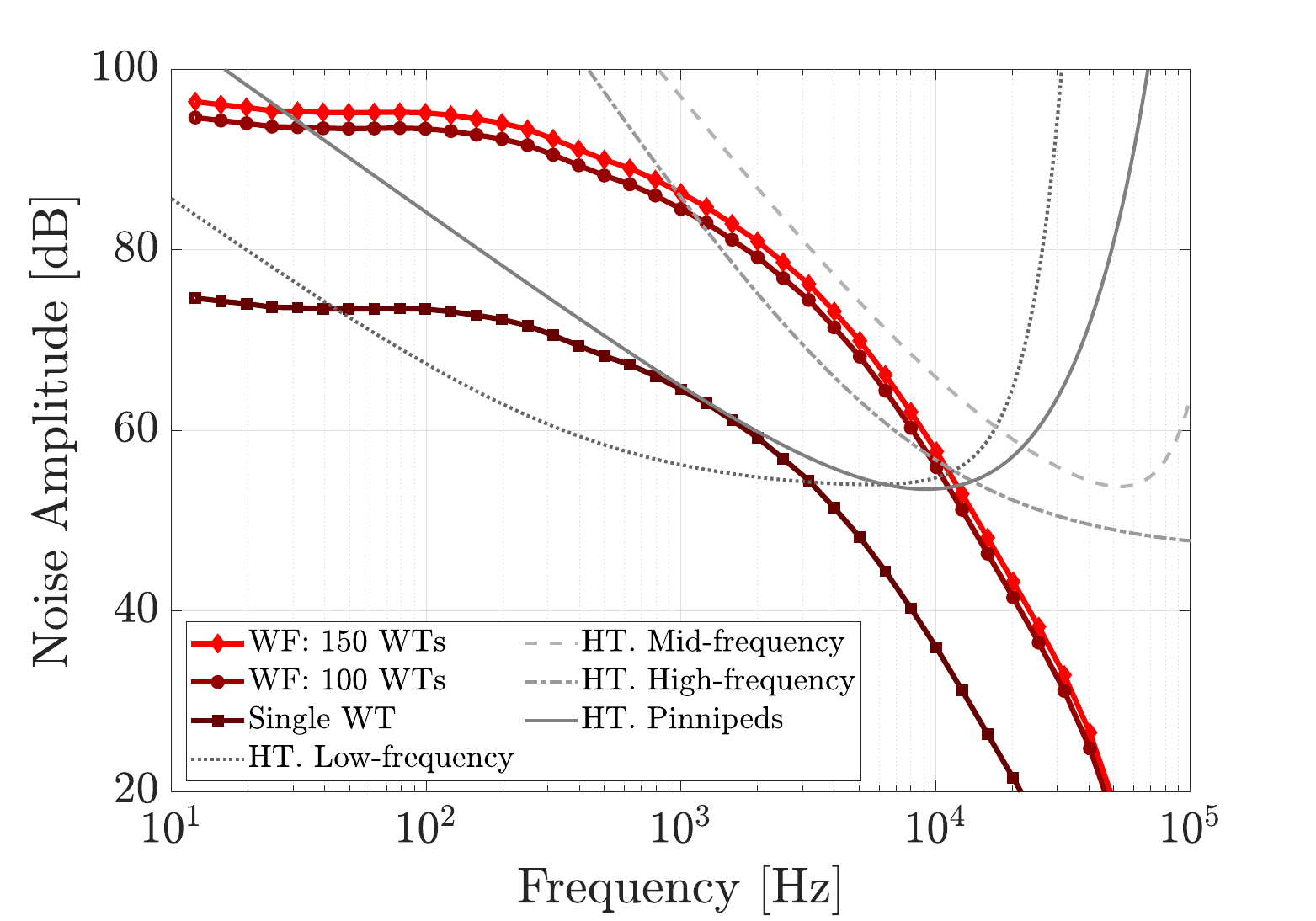}
		\vspace{0mm}\caption{IEA 22~MW wind turbine}
		\label{subfig:22MW}
	\end{subfigure}
        \hfill
	\begin{subfigure}[b]{0.49\textwidth}
		\centering
		\includegraphics[width=\textwidth]{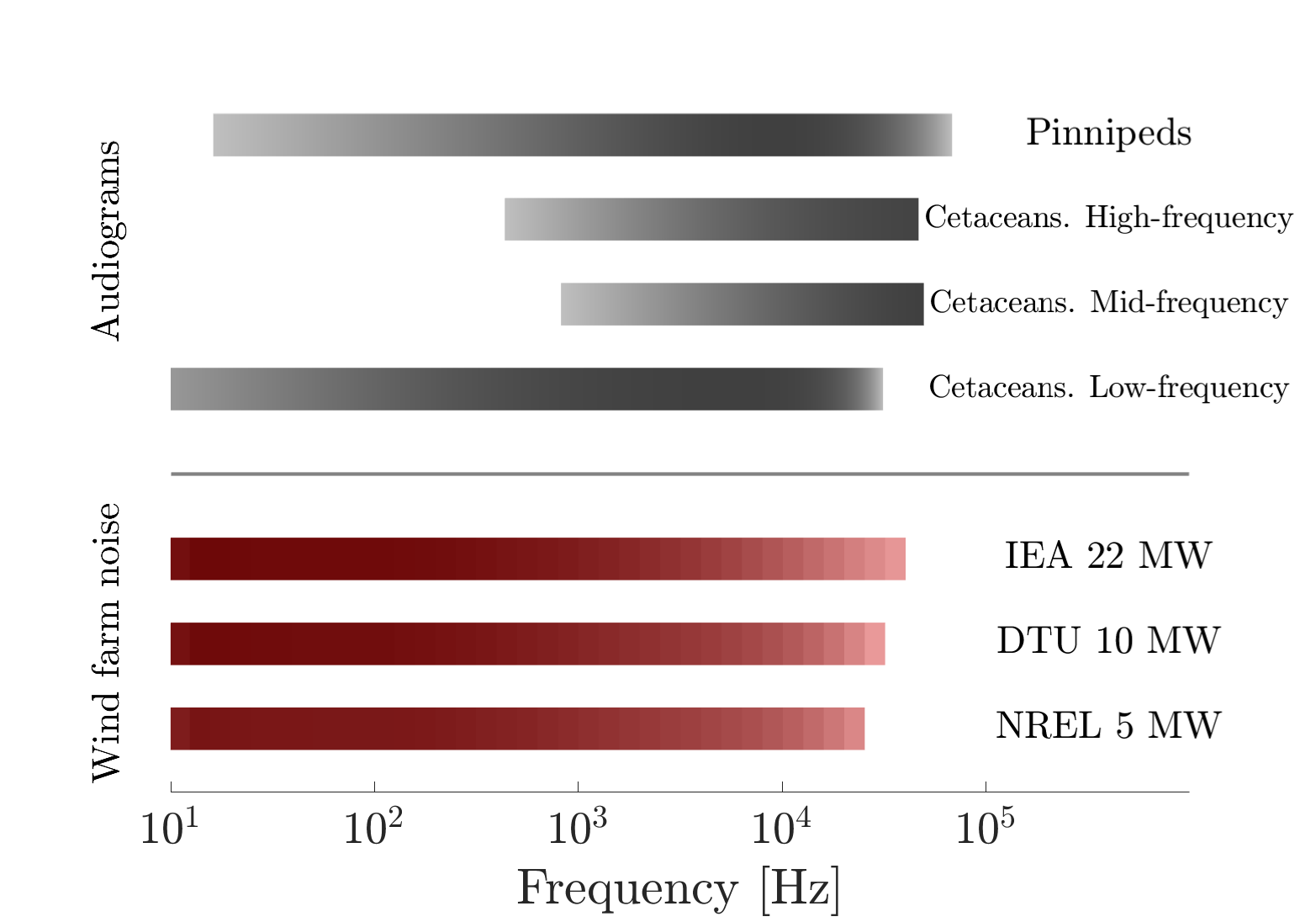}
		\caption{Overlap of wind farm noise (150 WT) and marine animals' audiograms}
		\label{subfig:all}
	\end{subfigure}
	\caption{One-third octave far-field noise spectra for a Water-side observer (10~m deep) at 100~m downwind the turbine compared to the hearing threshold of several groups of marine animals. 
    WT: wind turbine; WF: wind farm with 100 and 150 WTs; HT: hearing threshold. Color scales from 20-100 dB in figure (d).}
	\label{fig:ffn}
\end{figure}

\autoref{fig:ffn} shows the far-field aerodynamic noise spectra generated by a single wind turbine and groups of 100 and 150 wind turbines for each case, compared to the hearing thresholds of various functional hearing groups, i.e., low-, mid-, and high-frequency cetaceans and pinnipeds~\citep{Erbe2016,carter2013,tougaard2021}. The figure shows that aerodynamic noise from the three wind turbines affects the low-frequency hearing group even when considering only a single wind turbine.

\begin{figure}[h]
\centering
\includegraphics[width=0.5\textwidth]{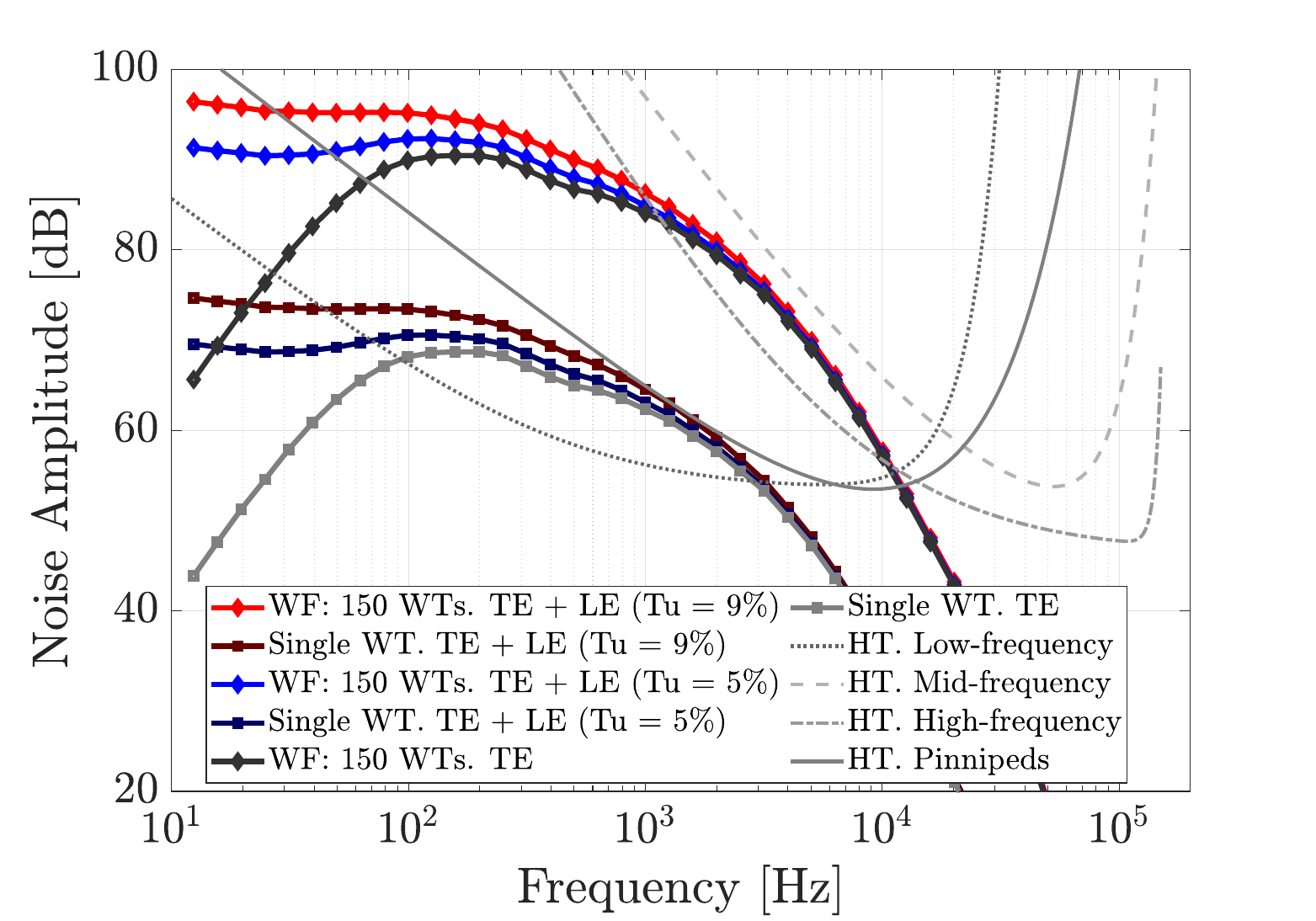}
\caption{Effect of the turbulence intensity on the aerodynamic noise generated by the IEA~22~MW wind turbine. Water-side observer  (10~m deep) at 100~m downwind of the turbine. TE: trailing-edge noise; LE: leading-edge noise; Tu: turbulence intensity}\label{fig:effect_turb}
\end{figure}

When considering farms,~\autoref{fig:ffn} shows that a group of 100 turbines causes a general increase in the amplitude of aerodynamic noise spectra, with a footprint underwater of 15~dB louder than the hearing threshold of some marine animals for the case of the 5~MW wind turbine and up to 25~dB for the case of the 22~MW wind turbine. In these cases, the aerodynamic noise of large offshore farms is much larger than the hearing threshold of several groups of marine species, which can potentially mask the sound naturally present in the environment. 
Having established a potential problem, we now focus on the largest turbine, the IEA~22~MW offshore turbine. The effect of inflow turbulence on the aerodynamic noise propagated underwater is shown in~\autoref{fig:effect_turb}. The same analyses have been conducted for the smaller turbines, showing similar results (not included for the sake of brevity). The results show that inflow turbulence noise significantly increases total aerodynamic noise in the low-frequency range (up to 1~kHz). This increase is perceived by low- and mid-frequency cetacean hearing groups. Low-frequency noise is more critical as a result of its longer propagating distance underwater (larger wavelengths), wich means that the acoustic footprint of the wind turbines is larger for lower frequencies. Therefore, turbulence intensity plays a role in the underwater aerodynamic noise of offshore wind turbines. However, the mechanical noise associated with the generator and radiated from the tower has a typical noise footprint for frequencies lower than 200~Hz and also needs to be considered to establish the dominance of LE noise of the total radiated noise of the wind turbine in this frequency range~\citep{madsen2006,Thomsen2006,Tougaard2009}.

Our study shows that the main source of noise from offshore wind turbines that affects marine animals is produced by the trailing edge of the blade. The latter is produced at frequencies where marine animals have their highest hearing sensitivity (near the inflection point of the audiograms). Furthermore, trailing-edge noise is generated for all operating conditions and does not depend on the inflow characteristics or mechanical components, which make this source of noise more critical. This information should drive manufacturers and regulators to push for the inclusion of noise reduction techniques in offshore wind turbines, such as trailing-edge serrations~\citep{oerlemans2016reduction}, and consider noise emissions in optimization-related tasks~\citep{defrutos2024deepreinforcementlearningmultiobjective}.

\begin{figure}[h]
\centering
\includegraphics[width=0.5\textwidth]{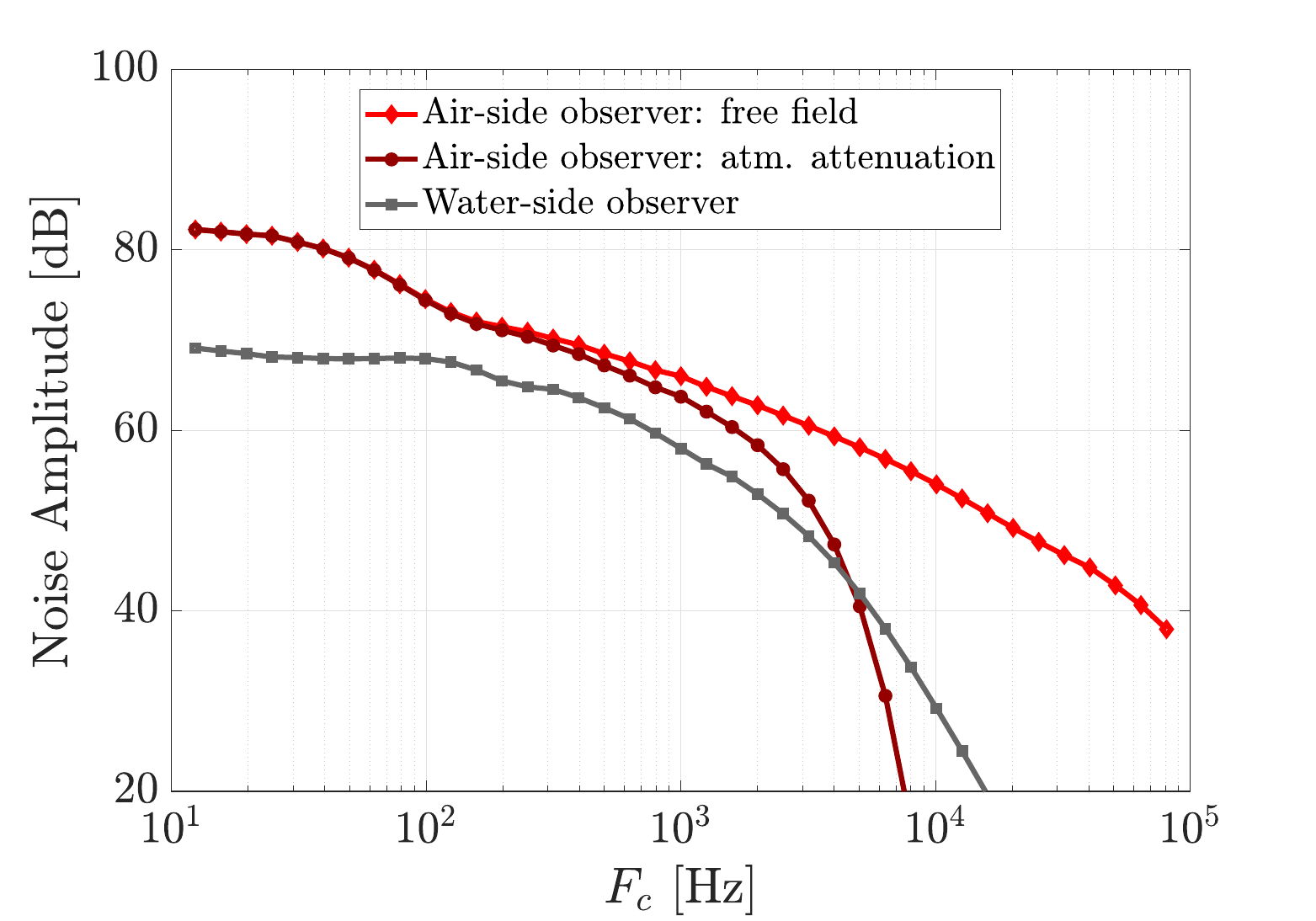}
\caption{IEA~22~MW wind turbine far-field noise prediction for a Water-side observer (10~m deep) and an Air-side observer at 500~m downwind of the turbine including (atm. attenuation) and neglecting (free field) atmospheric attenuation.}\label{fig:comp_all_cases}
\end{figure}

\autoref{fig:comp_all_cases} compares the far-field noise for an Air-side observer (i.e., located on the air-water surface) and for a Water-side observer at 10~m depth (see~\autoref{fig:noise_prop}), both located 500~m downwind of the turbine. For the former, the noise is computed directly at the observer location for two conditions: including and neglecting atmosphere attenuation. The figure shows that atmospheric attenuation causes a significant drop in far-field noise for frequencies higher than 1~kHz. If we also consider human hearing capacity (up to 20~kHz), then the aerodynamic noise of wind turbines is usually neglected at high frequencies (higher than 10~kHz). However, when considering underwater noise, the high-frequency range is less attenuated, and at the Water-side observer, the noise is louder than at the Air-side observer for frequencies higher than 5~kHz. This can be explained by direct airborne and indirect waterborne noise trajectories (see \autoref{sec:air_water_trans}). For a Water-side observer, the sound wave travels through the air (medium with the highest attenuation) a maximum distance of $\approx (H + R)/\cos(13^o)$, where $H$ is the height of the hub and $R$ is the wind turbine radius, which is much shorter than the distance from the source of the noise to an observer located downwind outside the water. This significantly reduces the atmospheric attenuation of the noise, which is more effective at higher frequencies. Furthermore, marine animals have hearing thresholds in a wider frequency range than humans, therefore, high frequencies are still relevant when analyzing marine environments.

\begin{figure}[ht!]
	\centering
		\begin{subfigure}[b]{0.49\textwidth}
		\centering
		\includegraphics[height=4.3cm]{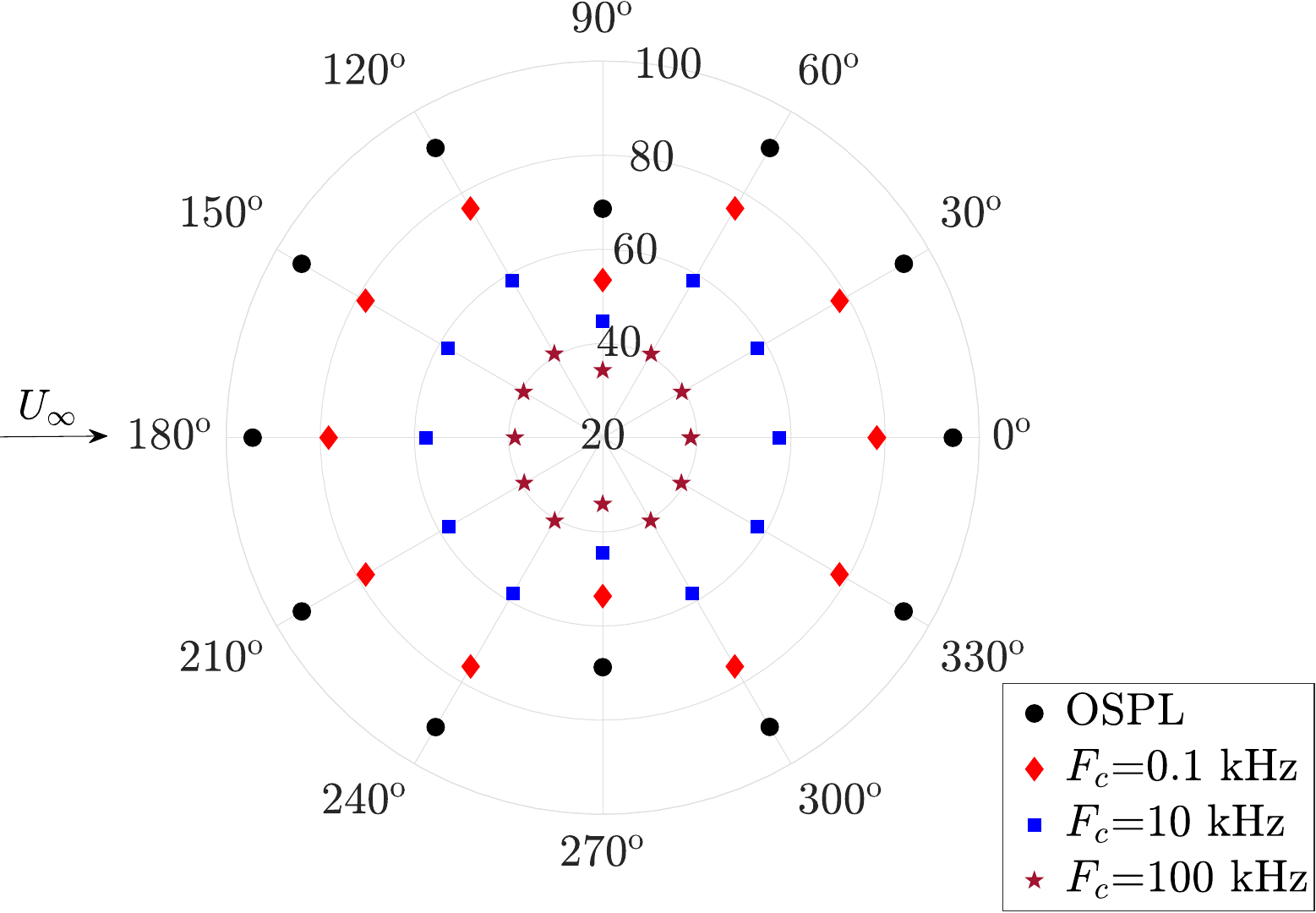}
		\caption{Air-side directivity}
		\label{subfig:dir_air}
	\end{subfigure}
	\begin{subfigure}[b]{0.49\textwidth}
		\centering
		\includegraphics[height=4.3cm]{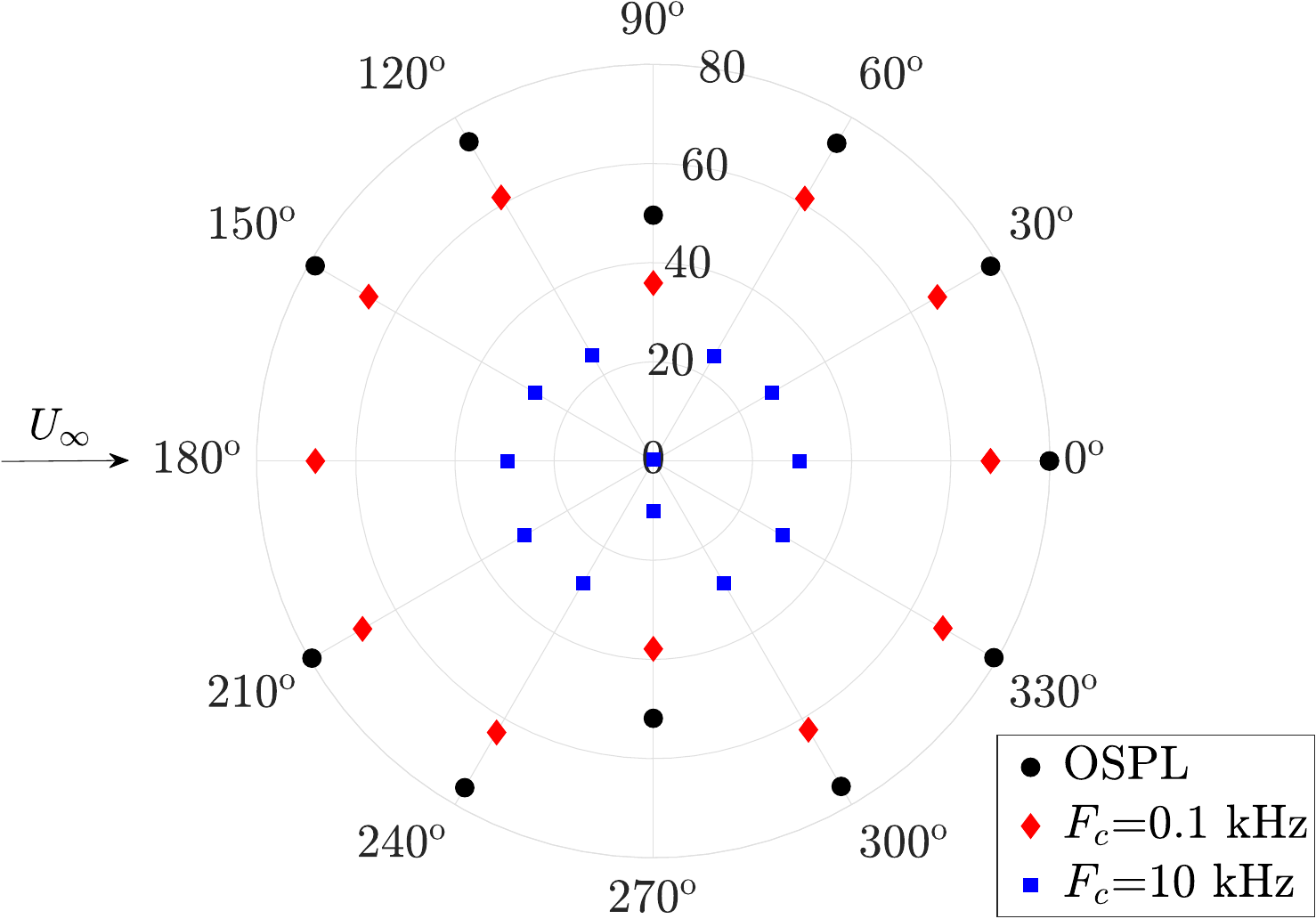}
		\caption{Water-side directivity}
		\label{subfig:dir_water}
	\end{subfigure}
	\caption{Direcitivity pattern of a single IEA~22~MW wind turbine for Air-side and 10~m deep Water-side observers 
 for several center frequencies and the overall sound pressure level (OSPL).}
	\label{fig:directivity}
\end{figure}

Finally, we discuss the directivity pattern in the air and underwater. \autoref{subfig:dir_air} shows the directivity radiation pattern of a single IEA~22~MW wind turbine considering Air-side observers at a radius of 500~m for several centered frequencies ($F_C$) and the integrated overall sound pressure level (OSPL) over the entire frequency range. In air, the wind turbine radiates noise as a dipole aligned with the wind inflow with the lowest radiation amplitude on the rotor plane. Consequently, the selected Air-side observer in~\autoref{fig:comp_all_cases} is aligned with the dipole main axis, whereas for the Water-side observer, only the noise that propagates underwater is the one near the rotor plane, following Snell's law. This also explains the lower noise radiated underwater than in air, mainly in the low-frequency range. 

\autoref{subfig:dir_water} shows the directivity of the IEA~22~MW wind turbine at 10~m depth. Underwater, the turbine also radiates noise as a dipole aligned with the inflow velocity, as in air. However, the dipole is not symmetric with respect to the inflow axis. The radiated noise is larger within the lower half of the figure (radiating more noise in the half-plane corresponding to 180 to 360\textsuperscript{o}). This region with higher noise corresponds to the blades rotating downward, which radiate higher noise~\citep{Oerlemans2007}. This asymmetry is not shown in air because the noise is propagated directly to the observer (and not following the Snell cone), and therefore the symmetry in directivity is preserved.

\conclusions \label{sec4}  
Environmental impacts of offshore energy devices must be carefully considered to achieve truly sustainable energy exploitation. A major concern with offshore wind turbines is underwater noise, which can negatively affect marine species. So far, turbine design for offshore environments has prioritized energy production, often neglecting its acoustic footprint. In particular, the aerodynamic noise transmitted underwater remains unquantified.

In this study, we have presented a methodology for predicting underwater aerodynamic noise from wind turbines by integrating existing noise prediction models with wave propagation theory. Such a methodology is a breakthrough to manufacturers and policymakers to quantify the environmental impact of wind energy devices related to noise emissions. The methodology shows a straight implementation for low-fidelity models used in optimization-related techniques or regulations. However, it can be easily coupled with high-fidelity methods to characterize the noise source (flow) or noise propagation. 

For the first time, we have quantified the underwater acoustic footprint of three large offshore turbines—$5$ MW, $10$ MW, and $22$ MW. We have compared the radiated noise with marine species' audiograms and scale the emissions for wind farms of 100 and 150 turbines. Our findings confirm that aerodynamic noise presents a potential environmental challenge for large offshore wind farms.

Specifically, we have identified the trailing-edge noise as the dominant aerodynamic source affecting marine environments. Despite significant attenuation at the air-water interface, this noise propagates underwater and remains perceptible to various marine species across a broad frequency range. To mitigate its impact, offshore wind turbine manufacturers must adopt noise reduction strategies already used in onshore turbines, such as serrated trailing edges. Addressing this issue will be crucial to balancing renewable energy expansion with the preservation of marine ecosystems.




\codedataavailability{The data that support the findings of this study are available from the corresponding author, Botero-Bolívar, L., upon reasonable request. Furthermore, the algorithm to predict wind turbine noise is available in~\citet{boteroBolivar2023_code}.} 



\appendix
\section{Algorithm of wind turbine noise prediction at the Water-side observer}\label{app:algoritmo}
\begin{figure}[h]
\centering
\includegraphics[width=0.85\textwidth]{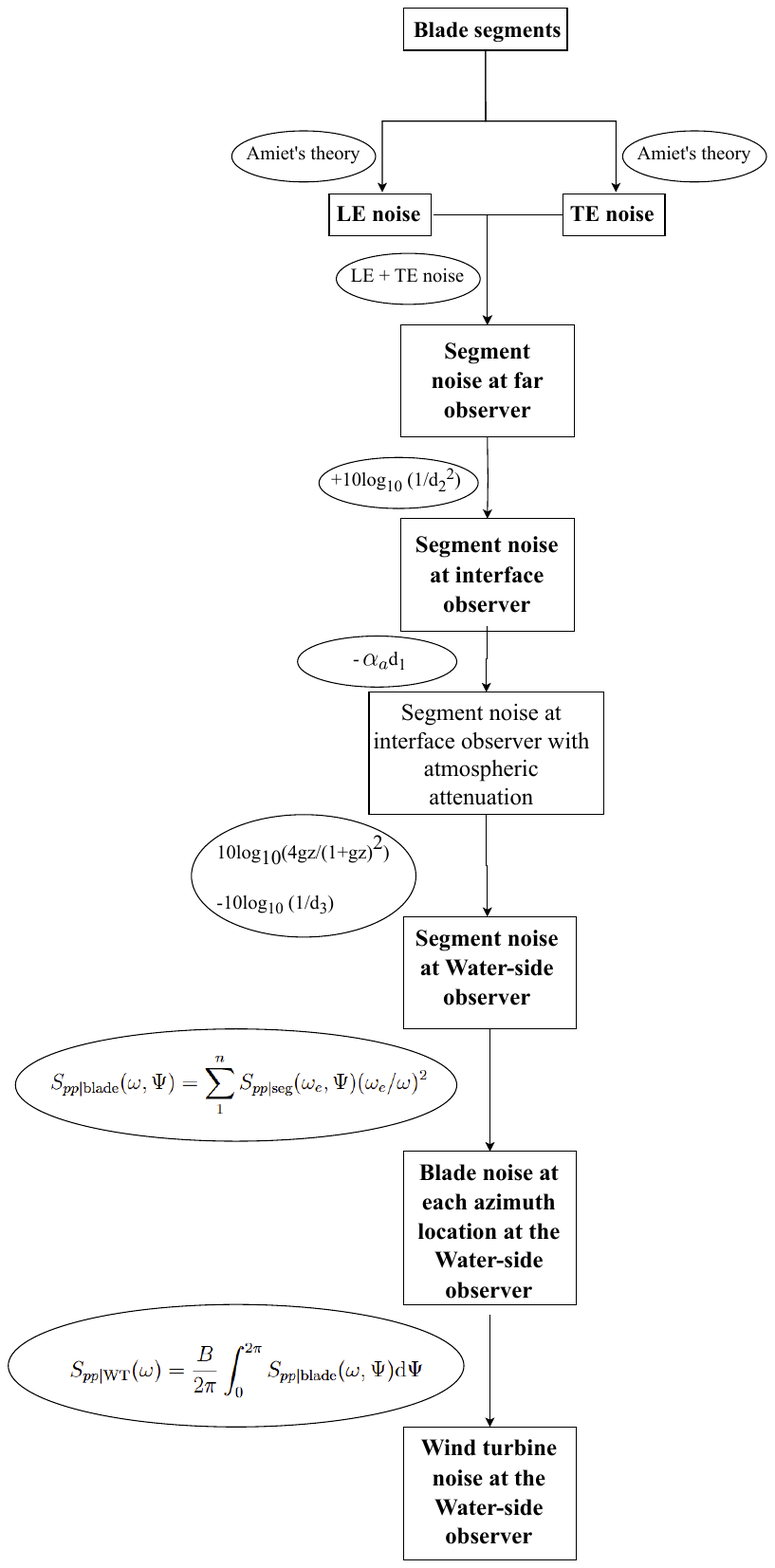}
\vspace{5mm}
\caption{Algorithm for the wind turbine noise prediction at the Water-side observer}\label{fig:algoritmo}
\end{figure}

\autoref{fig:algoritmo} shows the algorithm to calculate the aerodynamic wind turbine noise at the Water-side observer.
\appendixfigures

\noappendix       





\appendixtables   


\authorcontribution{\begin{itemize}
    \item Laura Botero-Bolívar: Conceptualization, Data curation, Formal analysis, Investigation, Methodology, Software, Validation, Visualization, Writing - original draft. 
    \item Oscar A. Marino: Investigation, Methodology,  Validation, Writing - review \& editing.
    \item Martín de Frutos: Investigation, Methodology,  Validation, Writing - review \& editing.
    \item Esteban Ferrer: Conceptualization, Methodology, Validation, Funding acquisition,  Project administration, Resources, Supervision, Writing - review \& editing.
\end{itemize}} 

\competinginterests{The authors have no relevant financial or non-financial interests to disclose.} 


\begin{acknowledgements}
This research has received funding from the European Union (ERC, Off-coustics, project number 101086075). Views and opinions expressed are, however, those of the authors only and do not necessarily reflect those of the European Union or the European Research Council. Neither the European Union nor the granting authority can be held responsible for them.
\end{acknowledgements}







\bibliographystyle{copernicus}
\bibliography{references.bib}

\end{document}